\documentclass[reprint,article,showpacs,preprintnumbers,amsmath,amssymb,citeautoscript,aip,superscriptaddress]{revtex4-1}
\usepackage{color,graphics}
\usepackage{graphicx}
\usepackage[sort&compress]{natbib}
\usepackage{natmove}
\usepackage{hhline}
\usepackage{soul}
\usepackage{xcolor}
\usepackage{array}
\usepackage{bm}
\usepackage{hyperref}

\usepackage[normalem]{ulem}
\usepackage[thinlines]{easytable}
\begin{document}

\setcitestyle{super}
\setcounter{totalnumber}{3}

\renewcommand{\thetable}{\arabic{table}}

\newcolumntype{P}[1]{>{\centering\arraybackslash}p{#1}}

\makeatletter

\newcommand{\thickhline}{%
    \noalign {\ifnum 0=`}\fi \hrule height 0.4pt
    \futurelet \reserved@a \@xhline}
\newcolumntype{"}{@{\hskip\tabcolsep\vrule width 1pt\hskip\tabcolsep}}
\makeatother

\title{Renormalization of excitonic properties by polar phonons}
%\title{Nonperturbative lattice effects on the excitonic properties of semiconductors}
%Nonperturbative exciton-phonon interactions in polar semiconductors}
%\title{Path integral approach for studying excitonic properties}

\author{Yoonjae Park}
 \affiliation{Department of Chemistry, University of California, Berkeley}
\author{David T. Limmer}
 \email{dlimmer@berkeley.edu}
 \affiliation{Department of Chemistry, University of California, Berkeley}
\affiliation{Materials Science Division, Lawrence Berkeley National Laboratory}
\affiliation{Chemical Science Division, Lawrence Berkeley National Laboratory}
\affiliation{Kavli Energy NanoScience Institute, Berkeley, California, Berkeley}

\date{\today}
\vspace{0mm}

\begin{abstract}
We employ quasiparticle path integral molecular dynamics to study how the excitonic properties of model semiconductors are altered by electron-phonon coupling. We describe ways within a path integral representation of the system to evaluate the renormalized mass, binding energy, and radiative recombination rate of excitons in the presence of a fluctuating lattice. To illustrate this approach, we consider Fr\"ohlich-type electron-phonon interactions and employ an imaginary time influence functional to incorporate phonon-induced effects \textcolor{black}{nonperturbatively}. The effective mass and binding energies are compared with perturbative and variational approaches, which provide qualitatively consistent trends. We evaluate electron-hole recombination rates as mediated through both trap-assisted and bimolecular processes, developing a consistent statistical mechanical approach valid in the reaction limited regime. These calculations demonstrate how phonons screen electron-hole interactions, generically reducing exciton binding energies and increasing their radiative lifetimes. 
\end{abstract}

\maketitle

%%%%%%%%%%%%%%%%%%%%%%%%%%%%%%%%%%%%%%%%%%%%%%%%%%%%%%%%%
%%%%%%%%%%%%%%%%%%%%%%%%%%%%%%%%%%%%%%%%%%%%%%%%%%%%%%%%%
\section{INTRODUCTION}

The application and design of photovoltaic devices rely on understanding the photophysics of semiconducting materials. Recent studies into novel low dimensional and hybrid perovskite semiconductors have highlighted the need to incorporate effects of a fluctuating lattice on the stationary behavior of excitons, moving away from the traditional perspectives in which screening is presumed to be largely determined by electronic degrees of freedom.\cite{umari2018infrared,thouin2019electron,miller2019tuning,liu2021exciton,schilcher2021significance,tao2022dynamic, tao2020dynamic} While the study of electron-phonon coupling for free charges has a long history, including foundational studies on polarons,\cite{frohlich2, slowelc, holstein, feynmanslow, polaronrev, ashok,sciadvzhu, ptemass} there is comparatively little known concerning the effects of electron-phonon coupling on excitonic properties. Motivated by observations that suggest polaronic effects play an important role in renormalizing exciton mobilities,\cite{reichman2018} binding energies, recombination rates \cite{herz2014, stranks2013} and photoluminescence yields \cite{nanolett.1c02122}, we aim to fill this knowledge gap. In this work, we explore \textcolor{black}{the effects of phonons} on the excitonic properties of traditional and hybrid perovskite materials using a path integral approach. Working within a Fr\"ohlich model Hamiltonian,\cite{frohlich2} we evaluate numerically exactly the role of phonons, finding that they generally reduce exciton binding energies and increase radiative lifetimes.

 The static properties of excitons determine the power conversion efficiencies of photovoltaics, the quantum yields for light emission and more. Predicting these properties from molecular models is an area of active development.\cite{ginsberg2020spatially} Most widely used approaches build upon ground state density functional theory, employing corrections from many-body physics including the GW approximation and Bethe-Salpeter equations.\cite{hedin1965new,albrecht1998excitonic,rohlfing2000electron} These and related approaches\cite{wang2020excitons} have been successful for a wide range of semiconducting materials \cite{giustino2014, malone2013} However, these theories traditionally ignore dynamical effects from phonons, as including them \textcolor{black}{within this framework beyond} is challenging. While historically, analytical approaches based on model Hamiltonians have been developed,\cite{haken1956quantentheorie,pbtheory,adamowski1981treatment} recent efforts have focused on numerical methods to describe the effects of phonons approximately.\cite{neatonPRL, GPRL2019,giustinoPRB,multiphonon}  While these approaches leverage powerful \emph{ab initio} many body theories, they have been limited in the strength of the electron phonon coupling that can be considered and the time and lengthscales approachable. 
 
Here, we present a path integral approach for describing electron and hole quasiparticles interacting with phonons. Analogous quasiparticle path integral approaches have been utilized to describe photoinduced phase separation \cite{bischak2017origin,piapp2020,bischak2018tunable}, charge trapping \cite{jefferson2020,chandler1994excess}, and charge recombination \cite{park2022}, as well as confinement effects.\cite{shumway2004quantum} We can derive an imaginary time influence functional, allowing us to incorporate dynamical and quantum mechanical effects of phonons \textcolor{black}{within a harmonic approximation}. To sample the resultant theory, we apply path integral molecular dynamics (MD) and study how electron-phonon coupling renormalizes the band mass, exciton binding energy, and electron-hole recombination rate for Fr\"ohlich-type interactions between electron and hole quasiparticles and phonons. In the followings, we elaborate theoretical details of path integral approach applied to exciton with phonons, and present how to calculate each property within this framework followed by the discussion on the effects of phonons.

%%%%%%%%%%%%%%%%%%%%%%%%%%%%%%%%%%%%%%%%%%%%%%%%%%%%%%%%%
%%%%%%%%%%%%%%%%%%%%%%%%%%%%%%%%%%%%%%%%%%%%%%%%%%%%%%%%%
\section{Theory}

We consider a system composed of an electron and hole interacting with a field of phonons, whose effective Hamiltonian consists of three parts,
\begin{equation}
\begin{aligned}
\hat{\mathcal{H}}_{\mathrm{}} = \color{black}{\hat{\mathcal{H}}_{\mathrm{eh}}} 
\color{black}{+ \hat{\mathcal{H}}_{\mathrm{ph}} + \hat{\mathcal{H}}_{\mathrm{int}}}
\end{aligned}
\label{Hall}
\end{equation}
a part due to the electronic degrees of freedom, \textcolor{black}{$\hat{\mathcal{H}}_{\mathrm{eh}}$}, a part from the lattice, $\hat{\mathcal{H}}_{\mathrm{ph}} $, and their interaction, $\hat{\mathcal{H}}_{\mathrm{int}}$. The electronic part includes the kinetic energies of an electron and a hole and a Coulomb interaction
\begin{equation}
\begin{aligned}
\color{black}\hat{\mathcal{H}}_{\mathrm{eh}} 
\color{black}= \frac{\hat{\mathbf{p}}_e^2}{2m_e} + \frac{\hat{\mathbf{p}}_h^2}{2m_h} - \frac{e^2}{4\pi \varepsilon_r |\hat{\mathbf{x}}_{e} - \hat{\mathbf{x}}_{h}|}
\end{aligned}
\label{Hel}
\end{equation}
where $\hat{\mathbf{p}}$ and $\hat{\mathbf{x}}$ are momentum and position operators of a quantum particle, $\varepsilon_r$ is the dielectric constant in units of the vacuum permittivity $\varepsilon_0$, and the subscript $e$ and $h$ indicate electron and hole. The masses  $m_e$ and $m_h$ are taken as their corresponding band masses using an effective mass approximation. This simplification can be relaxed by parameterizing more elaborate kinetic energy functions with {\color{black}position dependent masses.}\cite{shumway2004quantum}

The lattice is described by a collection of harmonic modes
\begin{equation}
\begin{aligned}
\hat{\mathcal{H}}_{\mathrm{ph}} = \frac{1}{2} \sum_{\mathbf{k}} (\hat{\mathbf{p}}_{\mathbf{k}}^2 + \omega_{\mathbf{k}}^2 \hat{\mathbf{q}}_{\mathbf{k}}^2)
\end{aligned}
\label{Hph}
\end{equation}
where $\hat{\mathbf{p}}_\mathbf{k}$ and $\hat{\mathbf{q}}_\mathbf{k}$ are the mass weighted momentum and coordinate of a phonon at wave vector $\mathbf{k}$. Without loss of generality, we will take the frequency of the oscillators to be constant $\omega_{\mathbf{k}}=\omega$, and equal to the longitudinal optical mode. While previous work has illustrated the importance of including additional modes or their wave-vector dependence in specific materials,\cite{multiphonon,park2022} we neglect these effects {\color{black}here in order to benchmark the approach to a simplified model.} For generalizations, see App.~\ref{App1}. 

We adopt a Fr\"ohlich-type interaction \cite{frohlich1, frohlich2} between the charges and the phonons, where a charged particle interacts linearly with the polarization field produced by a lattice vibration,
\begin{equation}
\begin{aligned}
\hat{\mathcal{H}}_{\mathrm{int}} =  \sum_{\mathbf{k}} \ \hat{\mathbf{q}}_{\mathbf{k}} \ \frac{C_{e} e^{i {\mathbf{k}} \cdot \hat{\mathbf{x}}_e} - C_{h} e^{i {\mathbf{k}} \cdot \hat{\mathbf{x}}_h}}{\mathbf{k}}
\end{aligned}
\label{Hint}
\end{equation}
where $\hat{\mathbf{q}}_{\mathbf{k}}$ corresponds to the polar displacement field that the charge can be coupled to along the $\mathbf{k}$ direction. The strength of the coupling is set by a material specific constant
\begin{equation}
\begin{aligned}
C_{i} = -i\hbar \omega 
\left( \frac{4\pi \alpha_i}{V} \right)^{\frac{1}{2}}
\left( \frac{\hbar}{2m_{i}\omega} \right)^{\frac{1}{4}}
\left( \frac{2\omega}{\hbar} \right)^{\frac{1}{2}}
\end{aligned}
\label{Calpha}
\end{equation}
where $i$ indicates either electron or hole, $V$ is the volume of the system, $\hbar$ is Plank's constant divided by $2\pi$ and $\alpha$ is a dimensionless Fr\"ohlich coupling constant \cite{frohlich1}. 
%QPIMD-Exciton-ph
To study this system we employ a path integral formalism \cite{feynman1,feynman2} which allows us to describe the correlated behavior of the electron, hole, and phonons quantum mechanically and \textcolor{black}{on an equal footing}.\cite{chandler1981exploiting} The partition function of the system can be written as
\begin{equation}
\begin{aligned}
\mathcal{Z}_{\mathrm{}} = \int \mathcal{D} [\mathbf{x}_e, \mathbf{x}_h, \mathbf{q}_{\mathbf{k}} ] \,
e^{-\mathcal{S} [\mathbf{x}_e, \mathbf{x}_h, \mathbf{q}_{\mathbf{k}} ]}
\end{aligned}
\end{equation}
where the path action $\mathcal{S}_{\mathrm{}}$ is defined as 
\begin{equation}
\begin{aligned}
\mathcal{S} = \frac{1}{\hbar} \int_{\tau=0}^{\beta \hbar} 
\color{black} \mathcal{H}_{\mathrm{eh,\tau}} 
\color{black} + \mathcal{H}_{\mathrm{ph,\tau}} + \mathcal{H}_{\mathrm{int,\tau}} 
\end{aligned}
\end{equation}
with the imaginary time variable $\tau$, and $\beta^{-1} = k_{\mathrm{B}}T$. The Hamiltonian indexed by $\tau$ represents the classical counterpart of Eqs.~\ref{Hel}-\ref{Hint} at given $\tau$ where $\hat{\mathbf{p}}$ and $\hat{\mathbf{x}}$ are replaced by $\mathbf{p}_{\tau}$ and $\mathbf{x}_{\tau}$ for each quasiparticle.

Considering that the phonons act as a Gaussian field coupled linearly to the charge density, the phonon variables $\{ \mathbf{q}_{\mathbf{k} }\}$ can be integrated out \cite{park2022, feynman1}, yielding
\begin{equation}
\begin{aligned}
\mathcal{Z}_{\mathrm{}} = \mathcal{Z}_{\mathrm{ph}} \int \mathcal{D} [\mathbf{x}_e, \mathbf{x}_h] 
\exp \left[ -\frac{1}{\hbar} \int_{\tau=0}^{\beta \hbar} 
\color{black}\mathcal{H}_{\mathrm{eh,\tau}} 
\color{black} + \mathcal{H}_{\mathrm{eff,\tau}}\right]
\end{aligned}
\end{equation}
with $\mathcal{Z}_{\mathrm{ph}}$ the partition function of phonons without the charge. The resulting effective Hamiltonian at a given $\tau$ can be written as a sum of four pieces
\begin{equation}
\begin{aligned}
\mathcal{H}_{\mathrm{eff,\tau}} = \sum_{i,j \in \{e,h\}} \mathcal{H}_{\mathrm{eff,\tau}}^{ij}
\end{aligned}
\end{equation}
with
\begin{equation}
\begin{aligned}
\mathcal{H}_{\mathrm{eff,\tau}}^{ij} = - \sigma_{ij} \frac{\alpha_{ij} \omega^2 \sqrt{\hbar}}{\beta \sqrt{8 m_{ij} \omega}} \int_{\tau' = 0}^{\beta \hbar } \frac{e^{-\omega |\tau - \tau'|}}{|\mathbf{x}_{i,\tau} - \mathbf{x}_{j,\tau'}|}
\end{aligned}
\end{equation}
where $\alpha_{ij} = \sqrt{\alpha_i \alpha_j}$, {$m_{ij} = \sqrt{m_i m_j}$}, and $\sigma_{ij}$ is $+1$/$-1$ for the same/opposite charges. We have used the inverse Fourier representation of $\mathbf{k}^{-2}$ with the equation for $C_{i/j}$ given in Eq.~\ref{Calpha}.
In order to use these results computationally, we discretized the imaginary time interval into $n$ slices. In this discrete formulation, the effective Hamiltonian $\mathcal{H} = \color{black}\mathcal{H}_{\mathrm{eh}} \color{black}+ \mathcal{H}_{\mathrm{eff}}$ becomes 
\begin{equation}
\begin{aligned}
\color{black}\mathcal{H}_{\mathrm{eh}} \color{black}=& 
\sum_{i,t} \frac{m_i n}{2\beta^2\hbar ^2} (\mathbf{x}_{i,t} - \mathbf{x}_{i,t+1})^2 
- \sum_{t} \frac{e^2}{4\pi \varepsilon_r n |\mathbf{x}_{e,t}-\mathbf{x}_{h,t}|}
\end{aligned}
\label{HRP}
\end{equation}
where we denote the last sum $\mathcal{H}_\mathrm{C}$ and 
\begin{equation}
\begin{aligned}
\mathcal{H}_{\mathrm{eff}}=\sum_{i,j} \mathcal{H}_{\mathrm{eff}}^{ij} = -\sum_{i,j} \sum_{t,s}
 \sigma_{ij} \frac{\alpha_{ij} \beta \hbar^{5/2} \omega^2 }{n^2 \sqrt{8 m_{ij}\omega} }   
\frac{e^{-\frac{\beta \hbar \omega}{n} |t - s|}}{|\mathbf{x}_{i,t} - \mathbf{x}_{j,s}|}
\end{aligned}
\label{Heff}
\end{equation}
where $i,j \in \{e,h\}$ and $t,s \in [1,n]$. The number of timeslices, or beads, is a convergence parameter that needs to be taken large for accuracy.\cite{virial} 

\begin {figure}
\centering\includegraphics [width=7.0cm] {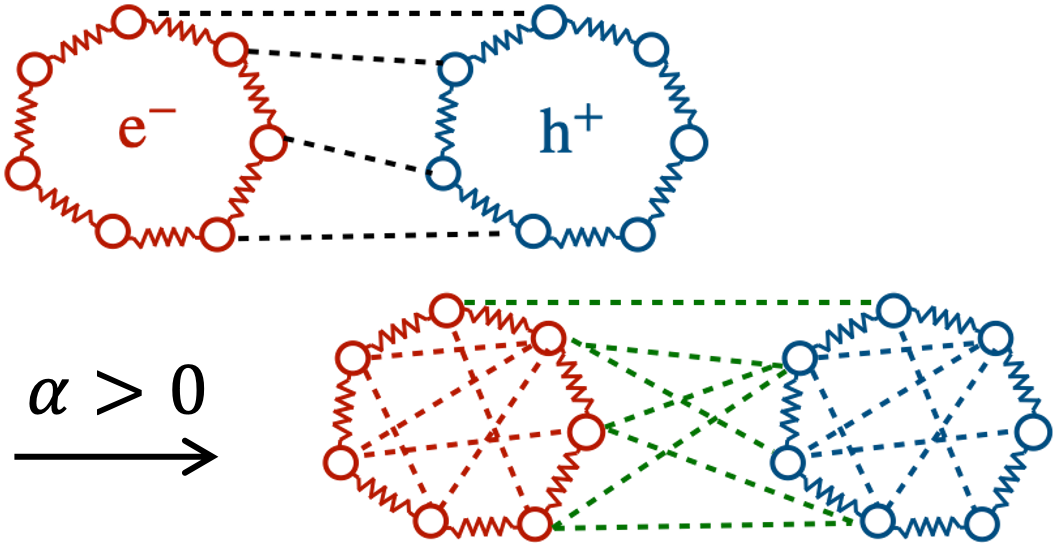}
\caption{\textcolor{black}{Illustration of imaginary time paths without (top) and with (bottom) phonon effects where black dotted lines represent a  bare Coulomb attraction, red and blue dotted lines describe the induced attractive self-interaction for the electron and hole, respectively, and the green dotted lines represent the effective screened interaction due to the dynamic phonons.}}
\label{IF}
\end{figure}

In the path integral framework, Eq.~\ref{HRP} implies that electron and hole quasiparticles are represented as classical ring polymers\cite{ceperley1995path} consisting of $n$ identical beads, where adjacent beads are harmonically coupled and beads with the same index from electron and hole are interacting through a fraction of Coulomb potential. Additionally from Eq.~\ref{Heff}, the effective energy induced by phonons depends on the positions of two different imaginary times, represented as the interaction between beads, and the stiffness of phonons sets the decaying imaginary timescale. In this way, we employ an imaginary time influence functional formalism which is shown schematically shown in Fig.~\ref{IF}.  Two effects from the phonons are clear from this picture. First, individual charges are localized by the phonons due to the induced attractive self-interaction. Second, the electron-hole interaction is weakened due to the induced repulsion, which is a reflection of phonon screening. The implications of these two effects are explored below. %The total error due to the discretization of this formalism scales by $n^{-2}$.

%%%%%%%%%%%%%%%%%%%%%%%%%%%%%%%%%%%%%%%%%%%%%%%%%%%%%%%%%
%%%%%%%%%%%%%%%%%%%%%%%%%%%%%%%%%%%%%%%%%%%%%%%%%%%%%%%%%

\section{Simulation details}

\renewcommand{\arraystretch}{1.5}
\begin{table}[b]
\centering
\begin{tabular}{P{4.7cm} | P{1.7cm} | P{1.7cm}}
\hline 
\hline 
Parameter (unit) & $ \textrm{CdS}$ & $\textrm{MAPbI}_3$  \\
\hline
electron band mass $m_e \, (m_0)$ & 0.19\cite{cdspara1} & 0.20\cite{mapipara1} \\
hole band mass $m_h \, (m_0)$ & 0.80\cite{cdspara1} & 0.20\cite{mapipara1} \\
optical frequency $\omega \, (\mathrm{THz})$ & 9.14\cite{cdspara2} & 7.53\cite{mapipara2} \\
dielectric constant $\varepsilon_{r} \, (\varepsilon_0)$ & 5.7\cite{cdspara1} & 6.1\cite{mapipara1} \\
band gap $E_{\mathrm{gap}} \, (\mathrm{eV})$ & 2.58\cite{cdspara1} & 1.64\cite{mapipara3} \\
Fr\"ohlich constant $\alpha_e$  (-) & 0.53\cite{ptemass} & 1.72\cite{mapipara2} \\
\hline 
\hline
\end{tabular}
\caption{Simulation parameters for CdS and MAPbI$_3$ \textcolor{black}{where $m_0$ is a bare mass of electron, $\varepsilon_0$ is the vacuum permmitivity, and the Fr\"ohlich constant for a hole can be calculated by $\alpha_h = \alpha_e \sqrt{m_h/m_e}$.}} 
\label{parameter}
\end{table}

To study the utility and efficiency of this approach, we consider models motivated by CdS and MAPbI$_3$, whose material properties are in different regimes of band mass and electron-phonon coupling strength. CdS is a traditional II-IV semiconductor where the material specific Fr\"ohlich coupling constant $\alpha$ is small, 0.53, and the band mass of the hole is much heavier than the mass of electron. For MAPbI$_3$, a lead-halide perovskite, the electron and hole have nearly the same band mass and the coupling strength is intermediate $\alpha =$ 1.72. For the purpose of this paper, since we aim to use the path integral method to study the general effects of electron-phonon coupling, we will ignore the anharmonic corrections from the lattice which can be important in determining the optoelectronic properties of these materials \cite{tao2022dynamic,herz2014, stranks2013,nanolett.1c02122, park2022}. Parameters used in simulations are summarized in Table \ref{parameter}. 

For both materials, we use MD simulations to sample the effective actions with fictious masses for the beads kept at 1 amu, and we study the renormalization of the effective mass, exciton binding energy, and recombination rate due to electron-phonon coupling.
For the effective mass calculations, $\mathcal{H}_{\mathrm{eff}}^{ee}$ defined in Eq. \ref{Heff} is used as the system Hamiltonian and for the other two properties, we run simulations of an electron-hole pair described as two ring polymers where one has a unit negative charge for an electron and the other has a unit positive charge for a hole with the Hamiltonian 
$\mathcal{H}_{\mathrm{}}= \color{black}\mathcal{H}_{\mathrm{eh}} \color{black}+ \mathcal{H}_{\mathrm{eff}}$ given in Eq.~\ref{HRP} and Eq.~\ref{Heff}. 

To avoid the divergence in the $1/|\mathbf{x}|$ term between attractive beads, a pseudopotential is used where $1/|\mathbf{x}|$ is replaced by $(r_{\mathrm{c}}^2+\mathbf{x}^2)^{-1/2}$ and $r_{\mathrm{c}}$ is chosen to reproduce the band gap of each material \cite{pseudo}. Simulations are run in an ensemble with constant volume, particle number and temperature using a Langevin thermostat  \textcolor{black}{where the total momentum averages to zero} with \textcolor{black}{integration time step} 1.0 fs and at room temperature unless explicitly specified, using the LAMMPS\cite{lammps} package.
For the following sections, we present a way to compute each property within the path integral framework followed by the discussion on the effects from phonons where $\alpha$ serves as a control parameter for the interaction strength between charges and phonons.

%%%%%%%%%%%%%%%%%%%%%%%%%%%%%%%%%%%%%%%%%%%%%%%%%%%%%%%%%
\section{Effective mass}

\begin {figure}
\centering\includegraphics [width=8.4cm] {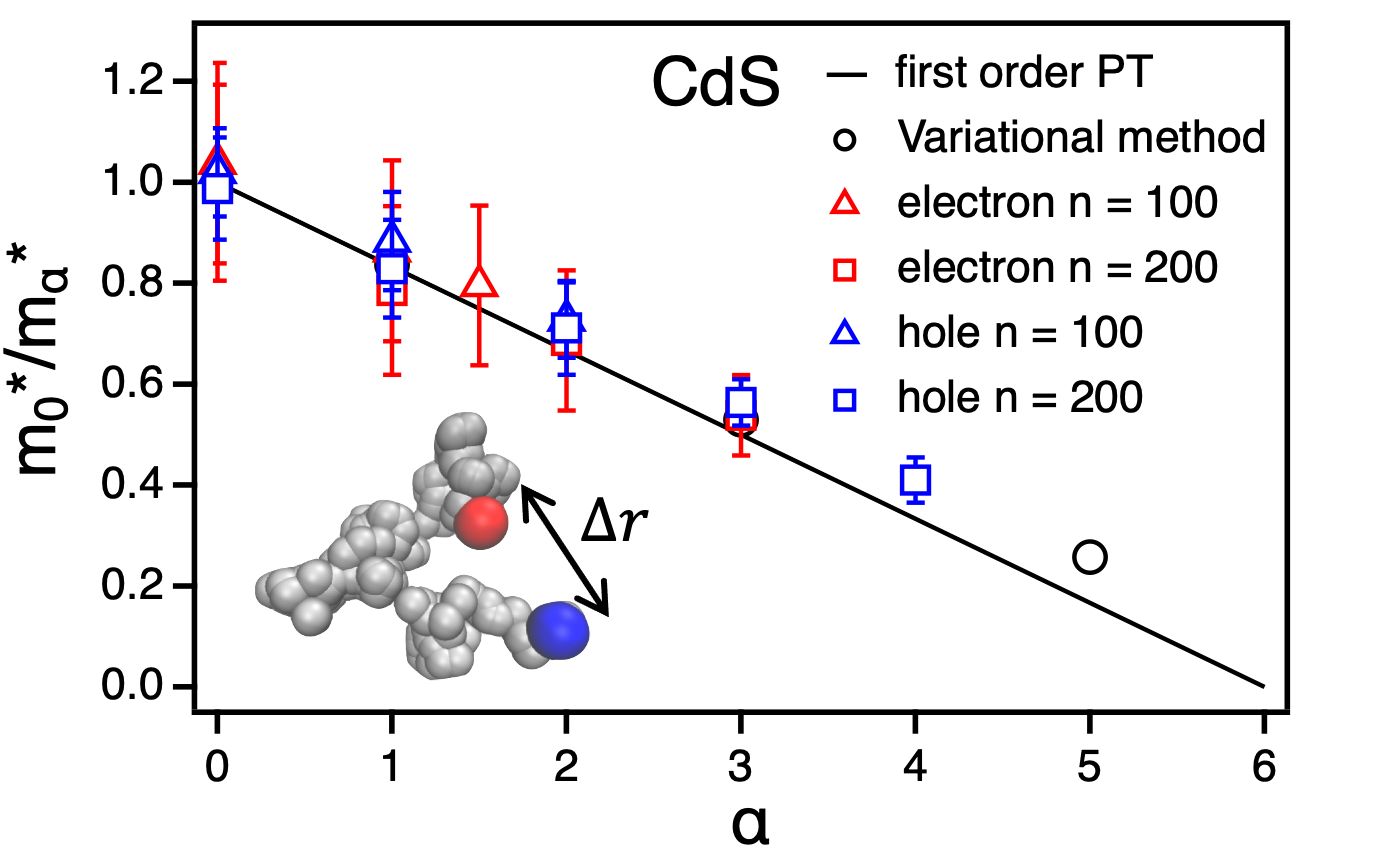}
\caption{Inverse effective mass with phonons $m^*_{\alpha}$ relative to the band mass $m^*_{0}$ as a function of coupling strength $\alpha$. Red/blue symbols represent the results for an electron/hole quasiparticles with different number of beads $n$. Black solid line and circles are the results predicted from first order perturbation theory and Feynman's variational approach \cite{Fvarm}. Inset figure is simulation snapshot for the schematic of $\Delta r$ where red and blue represent the first and the last beads.}
\label{emass}
\end{figure}

To validate the path integral framework, we first study the effective mass, for which significant previous analysis has been undertaken.\cite{feynman2,emass2} Since the presence of a charge induces a distortion to the lattice, motion of the charge requires moving the corresponding distortion field, renormalizes the mass of the charge, making it heavier.
Within the path integral framework, since the effective mass is a momentum-dependent quantity, 
\textcolor{black}{an inverse effective mass can be computed with an open-chain ring polymer in the low temperature limit as} \cite{emass1, emass2}
\begin{equation}
\begin{aligned}
\frac{1}{m^*} = \lim_{\beta \rightarrow \infty} \frac{\langle (\Delta r)^2 \rangle}{3\beta \hbar^2}
\end{aligned}
\label{invm}
\end{equation}
where $\langle .. \rangle$ denotes ensemble average and $\Delta r$ is the distance between the first and the last beads, schematically shown in Fig.~\ref{emass} inset. 

For simplicity, a Feynman unit system where $k_\mathrm{B}=\hbar = \omega = m = 1$ is used in this calculation. Given that Eq.~\ref{invm} is valid at low temperature, we tune the temperature and pseudopotential parameters and set $k_\mathrm{B} T/\hbar \omega=0.02$ and $r_{\mathrm{c}}=0.0707$ %% Units?
which are low and small enough for the convergence of our results.
Figure.~\ref{emass} describes the inverse effective mass of electron and hole quasiparticles of CdS at different $\alpha$ \textcolor{black}{where each point is the value averaged over 20 ensembles from simulations performed in constant volume, particle number, and temperature}. We find that the lighter mass and the stronger coupling require larger number of beads to converge.
The results from path integral approach are consistent with the value predicted from first order perturbation theory for small $\alpha$,\cite{ptemass}
\begin{equation}
\begin{aligned}
\frac{m^*_0}{m^*_{\alpha}} = 1 - \frac{\alpha}{6}
\end{aligned}
\end{equation}
shown in Fig.~\ref{emass}. The calculations are  consistent with Feynman's variational approach\cite{Fvarm} which is known as the most accurate solution across $\alpha$, demonstrating the validity of this framework and the $n$ required for convergence at various $\alpha$.

%%%%%%%%%%%%%%%%%%%%%%%%%%%%%%%%%%%%%%%%%%%%%%%%%%%%%%%%%
\section{Exciton binding energy}

Within the path integral framework, we capture the full correlation energy between the electron and hole quasiparticles, allowing us to compute accurate exciton binding energies. Here we show how electron-phonon coupling alters the binding energy. 
%where inclusion of electron-phonon coupling is controlled by coupling constant $\alpha$
We compute the exciton binding energy from the average energy of exciton \cite{binde} 
\begin{equation}
\begin{aligned}
\langle E \rangle = -\frac{\partial }{\partial \beta}
\ln \int \mathcal{D}[\mathbf{x}_e,\mathbf{x}_h] \, e^{-\beta \mathcal{H}_{\mathrm{}}}
\end{aligned}
\label{AvgE}
\end{equation}
resulting in two pieces, the average kinetic energy $\langle E \rangle_{\mathrm{K}}$ and the average potential energy $\langle E \rangle_{\mathrm{P}}$. For the kinetic energy, since the relevant terms produced by Eq.~\ref{AvgE} diverge as $n \rightarrow \infty$, we use a virial estimator \cite{virial}, which is known as an efficient way to estimate the kinetic energy in path integral simulations to avoid the large fluctuations from the subtraction of two diverging terms. Using the derivative of potential energy, the average kinetic energy can be written as
\begin{equation}
\begin{aligned}
\langle E \rangle_{\mathrm{K}} =& 3 k_{\mathrm{B}}T 
+ \frac{1}{2} \sum_{i \in \{e,h \}} \sum_{t=1}^n \left\langle  
\mathbf{x}_{i,t} \frac{\partial \left( \mathcal{H}_{\mathrm{C}} + 2\mathcal{H}_{\mathrm{eff}}^{eh} \right) }{\partial \mathbf{x}_{i,t}}\right\rangle \\
&+ \frac{1}{2}\sum_{t=1}^n \left\langle \mathbf{x}_{e,t} \frac{\partial \mathcal{H}_{\mathrm{eff}}^{ee}}{\partial \mathbf{x}_{e,t}}\right\rangle
+ \frac{1}{2}\sum_{t=1}^n \left\langle \mathbf{x}_{h,t} \frac{\partial \mathcal{H}_{\mathrm{eff}}^{hh}}{\partial \mathbf{x}_{h,t}}\right\rangle
\end{aligned}
\end{equation}
and the average potential energy becomes
\begin{equation}
\begin{aligned}
\langle E \rangle_{\mathrm{P}} 
= \langle \mathcal{H}_{\mathrm{C}} \rangle
+ \sum_{i,j \in \{e,h\}} 2\langle \mathcal{H}_{\mathrm{eff}}^{ij} \rangle
- \langle \mathcal{H}_{\mathrm{eff}}^{ij} \rangle'
\end{aligned}
\end{equation}
where $\langle \mathcal{H}_{\mathrm{eff}}^{ij} \rangle'$ is defined as $\langle \mathcal{H}_{\mathrm{eff}}^{ij} \rangle$ given by Eq.~\ref{Heff} with an additional factor of $\beta \hbar \omega |t-s|/n$ inside the summations with respect to $t$ and $s$.

The exciton binding energy is the energy threshold for an optical absorption between conduction and valence bands and thus traditionally reported in the low temperature limit. To evaluate it, we compute the average energy difference between the exciton and separately the electron and hole, at different temperatures and extrapolate to zero temperature, $E_{\mathrm{B}} = \lim_{T \rightarrow 0} \langle E \rangle_\mathrm{ex}-\langle E \rangle_{e}-\langle E \rangle_{h}$. \textcolor{black}{The subscript ex denotes an average with both electron and hole, while $e/h$ refers to calculations of the self energies of the electron/hole where the of quasiparticles interacts only with surrounding phonons.} 
%%
%\begin{equation}
%\begin{aligned}
%E_{\mathrm{B}} = \lim_{T \rightarrow 0} \langle E \rangle 
%\end{aligned}
%\end{equation}
%

\begin {figure}[t]
\centering\includegraphics [width=8.3cm] {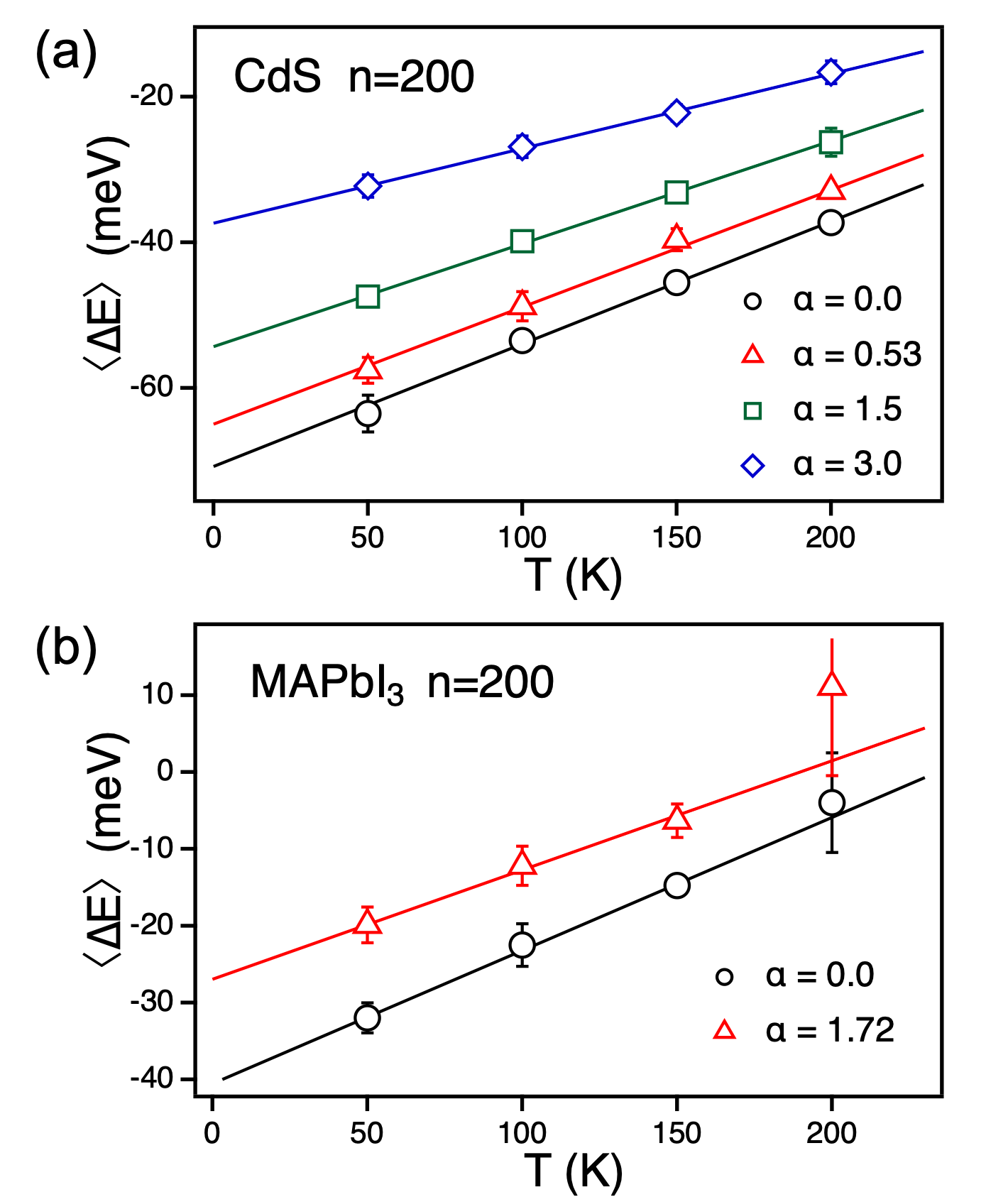}
\caption{Average energy difference computed from path integral simulations at different temperatures and coupling strength $\alpha$ for (a) CdS and (b) MAPbI$_3$ with 200 beads. Solid lines are linearly fitted lines.}
\label{bindE}
\end{figure}

Figures~\ref{bindE} (a) and (b) show the average energy difference at different temperatures for CdS and MAPbI$_3$. In both cases, we consider $\alpha=0$ and compare to $\alpha>0$, and find that $n=$ 200  is large enough to converge the result. The extrapolated exciton binding energies are summarized in Table \ref{EBsummary}. \textcolor{black}{We find that exciton energy becomes lower at high temperature, reflective of the higher population of phonons to stabilize the free charges.} Also, as the electron-phonon interaction becomes stronger, the exciton binding energy becomes smaller, implying that the phonons screen the effective electron-hole interaction. 

For the results without phonons, $\alpha = 0$, we compare our path integral simulations with binding energies from the Wannier-Mott exciton \cite{wannier} which is equivalent to our system.\cite{Hmodel} The exciton binding energy is computable exactly and given by $E_{\mathrm{B}}^H = \mu e^4 /2 (4\pi \varepsilon_r)^2 \hbar^2$ with $\mu=m_e m_h/(m_e+m_h)$ as the reduced mass of exciton. Calculated hydrogenic binding energies are 64.3 meV and 36.6 meV for CdS and MAPbI$_3$, consistent within a few meV with our values in both materials, showing the robustness of the calculation with the path integral framework. 
In the presence of electron-phonon interaction with nonzero value of $\alpha$, we compare the binding energies from the corresponding coupling strength, $\alpha=0.53$ for CdS and $\alpha=1.72$ for MAPbI$_3$, with the prediction from perturbation theory and Pollmann-Buttner theory \cite{pbtheory}.
The approximated differences in binding energies from the first order perturbation theory, $\Delta E_{\mathrm{B}}^F = E_{\mathrm{B},\alpha=0}-E_{\mathrm{B},\alpha \ne 0} \approx -2\alpha \hbar \omega$, are 6.4 meV and 17.1 meV for CdS and MAPbI$_3$, which are higher than our results, 5.8 meV for CdS and 13.6 meV for MAPbI$_3$.
Pollmann-Buttner theory results from a canonical transformation of the original Hamiltonian, which in the weak electron-phonon coupling limit provides an effective potential between the electron and hole.
Binding energies estimated by solving Schrodinger equation using Pollmann-Buttner potential between the electron and hole in App.~\ref{App2} are 41.3 meV and 16.7 meV for CdS and MAPbI$_3$, \textcolor{black}{lower} than results from the path integral simulations. 
We suspect the differences are attributed to the neglect of charge density relaxation due to hybridization with the phonons in perturbation theory and Pollmann-Buttner theory. This is likely a larger effect in MAPI$_3$ because of the equal masses of the electron and hole renders phonon screening at short distances a higher order process.  \textcolor{black}{Considering the fact that the anharmonicity from the lattice is not taken into account, the path integral estimates of the binding energies are in reasonable agreement with typical experimental values, of 28 meV\cite{toyozawa1986excitonic} and 16 meV\cite{herz2016} for CdS and MAPI$_3$.}

\renewcommand{\arraystretch}{1.53}
\begin{table}
\centering
\begin{tabular}{ P{2.4cm} | P{1.4cm} P{1.4cm} P{1.4cm} P{1.4cm} }
\hline 
\hline
CdS $\alpha$ & $ 0.0 $  &  0.53 & 1.5 & 3.0 \\
\hline
$E_\mathrm{B} \, (\mathrm{meV})$ & 70.8 & 65.0 & 54.3 & 37.4 \\
\hline
\thickhline
MAPbI$_3$ $\alpha$ & $ 0.0 $  &  1.72 & - & - \\
\hline
$E_\mathrm{B} \, (\mathrm{meV})$ & 40.6 & 27.0 & - & - \\
\hline
\hline 
\end{tabular}
\caption{Calculated exciton binding energy $E_{\mathrm{B}}$ for CdS and MAPbI$_3$ under different coupling strength.}
\label{EBsummary}
\end{table}

%%%%%%%%%%%%%%%%%%%%%%%%%%%%%%%%%%%%%%%%%%%%%%%%%%%%%%%%%
\section{Electron-hole recombination rate}

We now investigate the electron-hole recombination rate, which typically dominates the lifetime of charge carriers in bulk semiconducting materials.\cite{ginsberg2020spatially} 
It has been generally accepted that charge carrier recombination can be divided into three different mechanisms \cite{herz2016, herz2014}. The first is trap-assisted recombination where one charge carrier is trapped by a defect or impurity and then recombination occurs from the trap state. The second is due to bimolecular recombination where an electron in the conduction band is combined with a hole in the valence band, which is a predominant radiative pathway for direct semiconductors under standard operating conditions.\cite{Herz2016direct} Finally, Auger recombination in which electron and hole are recombined through an energy transfer to other charge carriers or phonons is yet a higher order process. While Auger recombination can be important in nanocrystals, it is typically negligible in bulk materials.\cite{philbin2018electron, philbin2020area}

All the mechanisms described above contribute to the recombination process but depending on the concentration of charge carriers, the dominant pathway varies. At low density, trap assisted recombination will dominate and at high density, Auger recombination will matter. In this work, we assume that the contribution from Auger recombination is small and consider the first and second order recombination processes. For a charge neutral system the electron and hole concentrations are equal $\rho_e= \rho_h$, and the rate equation is given by
\begin{equation}
\begin{aligned}
\frac{d \rho_e}{dt} = -k_{\mathrm{m}} \rho_e - k_{\mathrm{b}} \rho^2_e 
= -k_{\mathrm{tot}} \rho_e
\end{aligned}
\label{ratelaw}
\end{equation}
where  $k_{\mathrm{m/b}}$ is the rate constant for trap-assisted/bimolecular recombination process with $k_{\mathrm{tot}}$ as the overall rate constant. 

\begin {figure}
\centering\includegraphics [width=8.0cm] {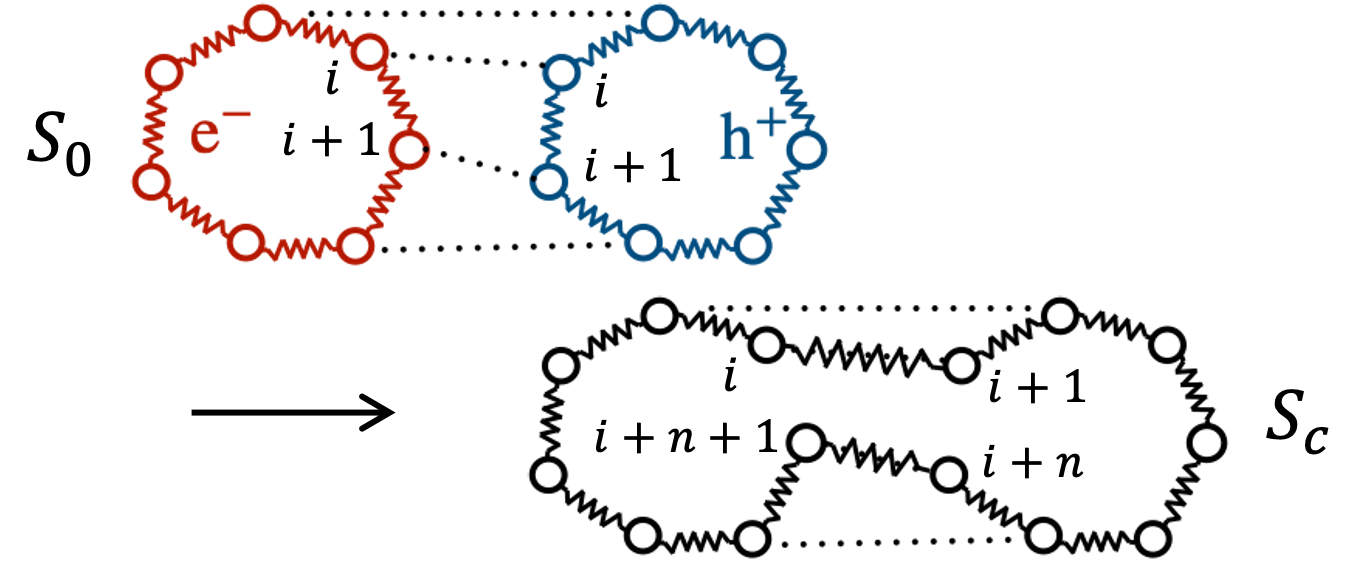}
\caption{Schematics of $\mathcal{S}_0$ and $\mathcal{S}_c$ where two separate paths (top) are combined at some imaginary time $i$ to form a combined radiating path (bottom).}
\label{deltas}
\end{figure}

Independent of the mechanism, the radiative recombination rate of an electron-hole pair within the path integral framework is given by \cite{shumway}
\begin{equation}
\begin{aligned}
k_{\mathrm{r}} = \frac{e^2 \sqrt{\color{black}\varepsilon_{r}} 
E_{\mathrm{gap}}^2}{2 \pi \varepsilon_0 \hbar^2 c^3 \mu } \,\frac{\mathcal{Z}_{c}}{\mathcal{Z}_{0}}
\end{aligned}
\label{ratekr}
\end{equation}
originating from Fermi's golden rule for spontaneous emission under our effective mass approximation, where $c$ is the speed of light and $E_{\mathrm{gap}}$ is the band gap energy. Other parameters for CdS and MAPbI$_3$ are summarized in Table \ref{parameter}.

The rate in Eq.~\ref{ratekr} is proportional to a ratio of partition functions whose subscripts 0 and $c$ indicate the standard thermal trace and the trace for a \emph{radiating path} where two separate imaginary time paths for quasiparticles are combined at a common imaginary time schematically shown in Fig.~\ref{deltas}. This ratio is identical to a thermally averaged overlap integral between the electron and hole densities. The partition functions are related through 
\begin{equation}
\begin{aligned}
\mathcal{Z}_c 
=\mathcal{Z}_0 \int \mathcal{D}[\mathbf{x}_e,\mathbf{x}_h] \, \frac{e^{- \mathcal{S}_{0}}}{\mathcal{Z}_0} e^{- \mathcal{S}_{c}+ \mathcal{S}_{0}}
=\mathcal{Z}_0 \left \langle e^{\Delta \mathcal{S}} \right \rangle_0
\end{aligned}
\end{equation}
where $\Delta \mathcal{S}$ is the change in path action. It is convenient for sampling purposes to rewrite this average using a conditional probability representation,
\begin{equation}
\begin{aligned}
P(\Delta \mathcal{S}) = \int d\mathbf{R} \, P(\Delta \mathcal{S}|\mathbf{R}) P(\mathbf{R})
\end{aligned}
\end{equation}
so that the ratio in the dilute limit can be written as 
\begin{equation}
\begin{aligned}
\frac{\mathcal{Z}_{c}}{\mathcal{Z}_{0}} =\rho_e \int d\mathbf{R} \,\left \langle e^{\Delta \mathcal{S}}\right \rangle _\mathbf{R} \ e^{-\beta \Delta F(\mathbf{R})}
\end{aligned}
\end{equation}
where $\mathbf{R}$ is the vector between electron and hole ring polymers, represented by $\beta F(R)=-\ln P(R)$ a free energy for changing their distance. The difference in path actions is equal to  
\begin{equation}
\begin{aligned}
\Delta \mathcal{S}
=& \sum_{t=1}^n \frac{m_e n}{2\beta\hbar ^2} (\mathbf{x}_{e,t} - \mathbf{x}_{e,t+1})^2 
+ \frac{m_hn}{2\beta\hbar ^2} (\mathbf{x}_{h,t} - \mathbf{x}_{h,t+1})^2 \\
&- \sum_{t=1}^{2 n} \frac{\mu n}{\beta\hbar ^2} (\mathbf{x}_{c,t} - \mathbf{x}_{c,t+1})^2
\end{aligned}
\end{equation}
for going between the thermal and radiating paths. 
The first two terms correspond to $\mathcal{S}_0$, and $\mathcal{S}_c$ is given by the last term, and the index of beads for the combined coordinates $\{ \mathbf{x}_c \}$ is schematically described in Fig.~\ref{deltas}. 
In the following, we present the details on simulations and discussions on trap-assisted and bimolecular recombination rates, and combine these two to estimate a total rate constant. In both, we assume that recombination is not limited by diffusion of the charge carriers, so that a local equilibrium distribution is established for the relative positions of electrons and holes. For both, we will evaluate the rate at room temperature $T=298 K$.

%%%%%%%%%%%%%%%%%%%%%%%%%%%%%%%%%%%%%%%%%%%%%%%%%%%%%%%%%
\subsection{Trap assisted recombination rate}

For trap-assisted recombination, we need to describe the trapping of a charge as well as the subsequent recombination of electron and hole quasiparticles. To describe this process we assume that the trapped charge achieves a steady state population, and the rate is given by the likelihood of finding a trapped charge times the rate to recombine that trapped charge with an incoming charge,
\begin{equation}
k_\mathrm{m} = \textcolor{black}{P_{\mathrm{trap}}} k_\mathrm{r}
\end{equation}
where \textcolor{black}{$P_{\mathrm{trap}}$ is the probability of an electron to be trapped.} 
%$\rho_\mathrm{trap}$ is the concentration of trapped charges. 
This approximation is valid in the dilute limit, provided trapping is reversible. For both CdS and MAPbI$_3$, we consider the trapping of an electron with a positively charged point defect. The point defect is described by a Coulomb potential acting between the defect and charge, determined by the corresponding dielectric constant $\varepsilon_r$ and $r_{\mathrm{c}}$ which is set to recover the reported trapping energy, 1.75\cite{trapcds}(0.3\cite{trapmapi}) eV for CdS (MAPbI$_3$).

\begin {figure}[b]
\centering\includegraphics [width=8.3cm] {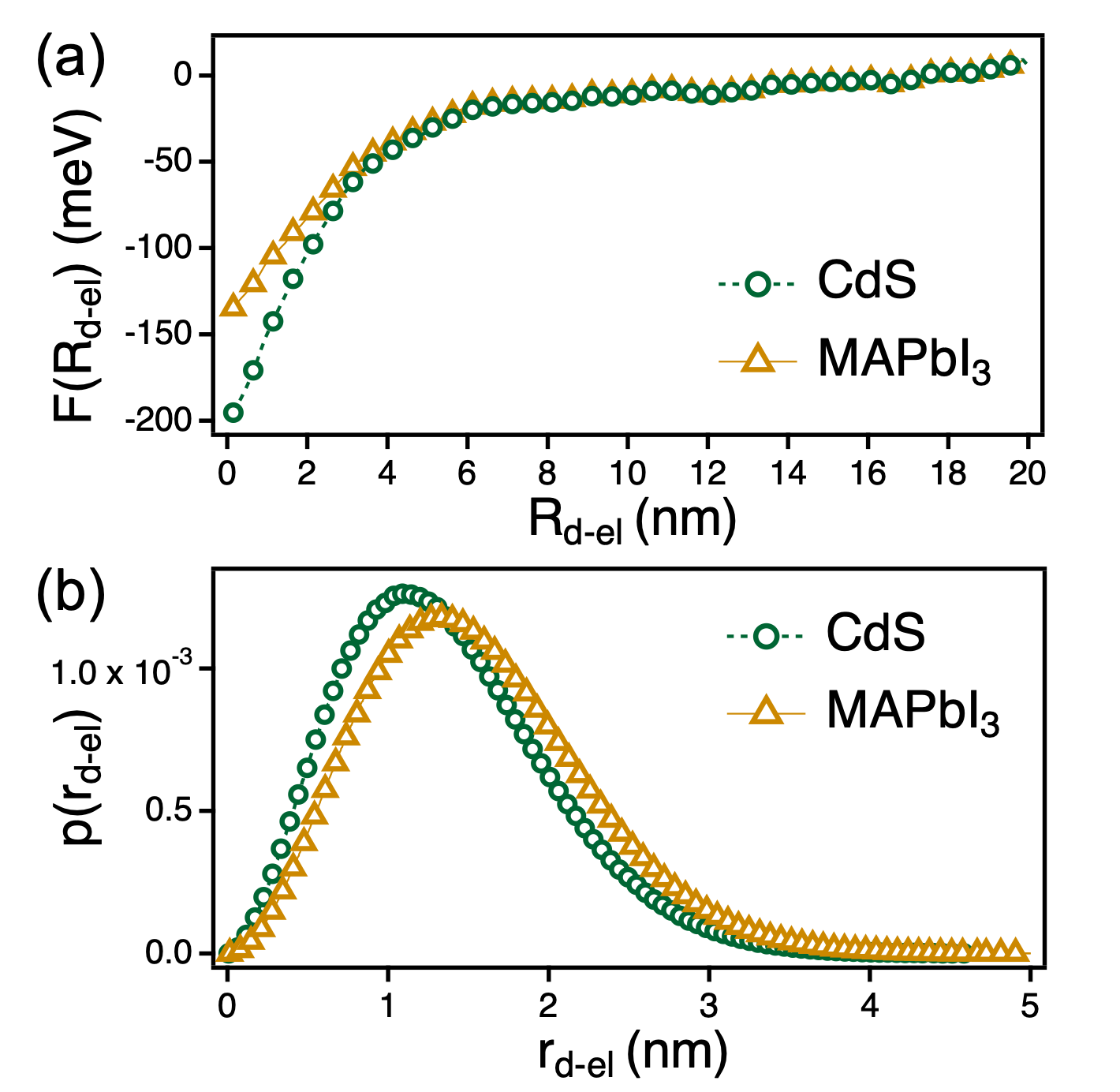}
\caption{(a) Free energy as a function of the distance between a defect and electron from 1 ring polymer simulation for CdS (green) and MAPbI$_3$ (orange). (b) Charge distribution for an electron localized on a defect for CdS (green) and MAPbI$_3$ (orange). Lines are guides to the eye.}
\label{monoFE}
\end{figure}

In equilibrium, the probability of finding an electron trapped by an isotropic point defect is given by
\begin{equation}
\begin{aligned}
\textcolor{black}{
P_{\mathrm{trap}} = \, 4\pi \rho_{\mathrm{d}} \int_0^{R_d^*} dR_d \, R_d^2  e^{-\beta \Delta F_{d}(R_d)}}
%
%\rho_{\mathrm{trap}} = \, 4\pi \rho_{\mathrm{d}}  \rho_{{e}} \int_0^{R_d^*} dR_d \, R_d^2  e^{-\beta \Delta F_{d}(R_d)}
\end{aligned}
\end{equation}
where $\rho_{\mathrm{d}}$ is the density of defect sites in the lattice where it is possible to trap an electron, $R_d^*$ is a cutoff distance for defining the trapped state, and $F_d(R_d)$ is the potential of mean force between the point defect and the electron, assuming both are dilute. To evaluate $P_\mathrm{trap}$, we employ umbrella sampling\cite{usamp} and Weighted Histogram Analysis Method \cite{wham1, wham2} by adding a bias potential $V(R_d) = 0.5 k_{\mathrm{sp}}(R_d - R_{\mathrm{eq}})^2$ on the distance between the defect and the centroid of the electron ring polymer with $k_{\mathrm{sp}}=0.2$\,kcal/mol/$\mathrm{\AA}^2$ and $\{ R_{\mathrm{eq}} \} = \{3\mathrm{\AA}, \ 6\mathrm{\AA}, \dots, 195\mathrm{\AA} \}$. This allows us to determine the potential of mean force,
\begin{equation}
\begin{aligned}
\beta F_d(R_d) = - \ln \langle \delta (R_d - |\mathbf{x}_{d}-\mathbf{x}_{e}^c|) \rangle 
\end{aligned}
\end{equation}
with $R_d$ as the distance between the defect $\mathbf{x}_{d}$ and the centroid of the electron ring polymer $\mathbf{x}_{e}^c$. 

The potential of mean force between an electron and the defect is shown in Fig.~\ref{monoFE} a). For both CdS and MAPbI$_3$, the potential is monotonic. The binding free energy of the electron to the defect is much less than the bare potential energy of the pseudo-potential, reflecting the charge delocalization. This delocalization is evident in Fig.~\ref{monoFE} b) which describes the charge distribution of the electron as a function of the distance between a point defect, $r_{d-el}=|\mathbf{x}_{d}-\mathbf{x}_{e}|$. For a free electron at large distance, the electron has a spatial extent defined by the radius of gyration of the imaginary time slices which is related to the thermal wavelength $\lambda$ at given temperature and mass, $R_g (\beta,m) = \hbar \sqrt{\beta}/ 2\sqrt{m}$. For CdS and MAPbI$_3$ these free particle sizes are both nearly 20  $\mathrm{\AA}$. Upon trapping to a defect, the electrons of both CdS and MAPbI$_3$ become slightly more localized, with characteristic sizes of 15  $\mathrm{\AA}$ and 17  $\mathrm{\AA}$ respectively.  

Given an equilibrium concentration of trapped electrons, the recombination rate is than evaluated by computing the likelihood of finding a hole in the vicinity of the electron and then measuring the conditional overlap in their densities, as described in Eq.~\ref{ratekr}. To evaluate both, we run MD simulations with a hole ring polymer as well as an electron ring polymer trapped into the point defect where Eq.~\ref{HRP} and \ref{Heff} are used for two ring polymers. To sample the trajectory efficiently, we add the same harmonic potentials described above along the distance between two centroids of ring polymers and additional harmonic potentials with $k_{\mathrm{sp}}=0.5$kcal/mol/$\mathrm{\AA}^2$ and $R_{\mathrm{eq}}=0.0$ on the distance between point defect and the centroid of electron ring polymer to hold an electron near the defect. This potential is unweighted analogously with WHAM to yield an unbiased distribution. 

The resultant trap-assisted rate constants under different electron-phonon coupling strengths with $\rho_{\mathrm{d}} = 10^{18} \mathrm{cm}^{-3}$  are summarized in \textcolor{black}{Table \ref{ratesummary}}. We find that the interaction with phonons reduces the rate constant in both materials \textcolor{black}{although values with finite coupling strength in CdS are not significantly distinct.} \textcolor{black}{This reduction results from the smaller likelihood of finding a hole in the vicinity of the electron,} a manifestation of dynamical screening from the phonons. This is explored more directly for bimolecular recombination below.

% MAPI mono.rate = 14 inverse $\mu$ s (Adv. Matter., Vol 26, 1584-1589, 2014)
% MAPI $\tau_m$ = $k_m^-1$ = 4ns ~ 1$\mu$ s
% CdSe/CdS core-shell Auger lifetime = 1 - 10 ns (Nano Lett. 2016, 16, 6491−6496)
% CdSe/CdS core-shell Auger lifetime = ~50 ns (ACS Nano, Vol 7, No. 4, 3411–3419, 2013) 
% Au/CdS core-shell lifetime = < 1 ns (ACS Nano, Vol 8, No. 1, 352-361, 2014)

\renewcommand{\arraystretch}{1.53}
\begin{table}
\centering
\begin{tabular}{ P{2.4cm} | P{1.4cm} P{1.4cm} P{1.4cm} P{1.4cm} }
\hline 
\hline
CdS $\alpha$ & $ 0.0 $  &  0.53 & 1.5 & 3.0 \\
\hline
$k_{\mathrm{m}} \, (\rm{\mu s}^{-1})$  & 0.93 & 0.37 & 0.36 & 0.44 \\
$k_{\mathrm{b}}\rho_e \, (\rm{ns}^{-1})$  & 0.101 & 0.082 & 0.032 & 0.025 \\
$\tau_{\mathrm{tot}} \, (\rm{ns})$  & 9.79 & 12.2 & 31.3 & 39.6 \\
\hline
\thickhline
MAPbI$_3$ $\alpha$ & $ 0.0 $  &  1.72 & - & - \\
\hline
$k_{\mathrm{m}} \, (\rm{\mu s}^{-1})$ & \textcolor{black}{0.115} & \textcolor{black}{0.073} & - & - \\
$k_{\mathrm{b}}\rho_e \, (\rm{ns}^{-1})$ & \textcolor{black}{0.067} & \textcolor{black}{0.028} & - & - \\
$\tau_{\mathrm{tot}} \, (\rm{ns})$ & \textcolor{black}{15.0} & \textcolor{black}{35.5} & - & - \\
\hline
\hline 
\end{tabular}
\caption{Calculated trap-assisted rate $k_{\mathrm{m}}$ and bimolecular recombination rate constant $k_{\mathrm{b}}$ with the defect density $\rho_{\mathrm{d}} = 10^{18} \mathrm{cm}^{-3}$ and the carrier density $\rho_e = 10^{17} \mathrm{cm}^{-3}$, and carrier lifetime $\tau_{\mathrm{tot}}$ from $k_{\mathrm{tot}}$ for CdS and MAPbI$_3$ under different coupling strength. The mean statistical errorbar in these estimates is 10\%.}
\label{ratesummary}
\end{table}
% CdS Ptrap = 0.1789
% MAPI Ptrap = 0.01776

%%%%%%%%%%%%%%%%%%%%%%%%%%%%%%%%%%%%%%%%%%%%%%%%%%%%%%%%%
\begin {figure}[t]
\centering\includegraphics [width=8.3cm] {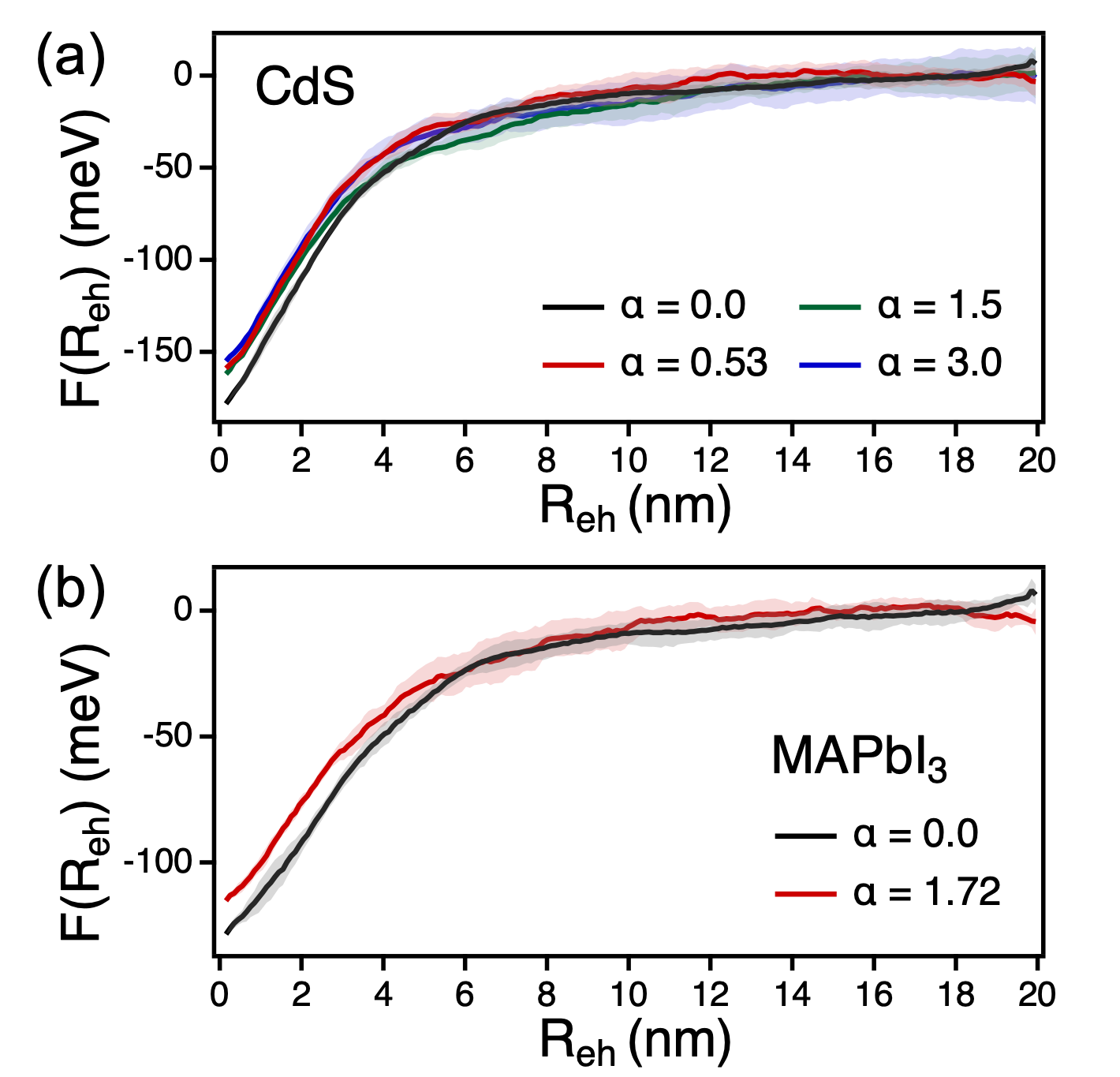}
\caption{Potentials of mean force for the charge centers. Potentials of mean force for (a) CdS and (b) MAPbI$_3$ for a variety of electron-phonon coupling strength $\alpha$. Error bars are shown as shaded regions.}
\label{biFE}
\end{figure}
\subsection{Bimolecular recombination rate}

Bimolecular recombination is studied through the same method described above with MD simulations of electron and hole ring polymers. The bimolecular recombination rate, $k_\mathrm{b} \rho_e=k_\mathrm{r}$, requires us to evaluate the potential of mean force for localizing the two charges and their subsequent overlap density. The potential of mean force, $F(R_{eh})$, between the centroids of electron $\mathbf{x}_e^c$ and hole $\mathbf{x}_h^c$ ring polymers,   
\begin{equation}
\begin{aligned}
\beta F(R_{eh}) = - \ln \langle \delta (R_{eh} - |\mathbf{x}_e^c-\mathbf{x}_h^c|) \rangle 
\end{aligned}
\end{equation}
is computable from umbrella sampling using the same procedure as that for the $R_d$. The resulting function is shown in Fig.~\ref{biFE} for both CdS and MAPbI$_3$ as functions of electron-phonon coupling, $\alpha$. The monotonic free energies display systematic destablization of the electron-hole pair with increasing $\alpha$, consistent with the reduction in exciton binding energies. Similarly, the minimum is more shallow for MAPbI$_3$ than for CdS. 

\begin {figure}[t]
\centering\includegraphics [width=8.3cm] {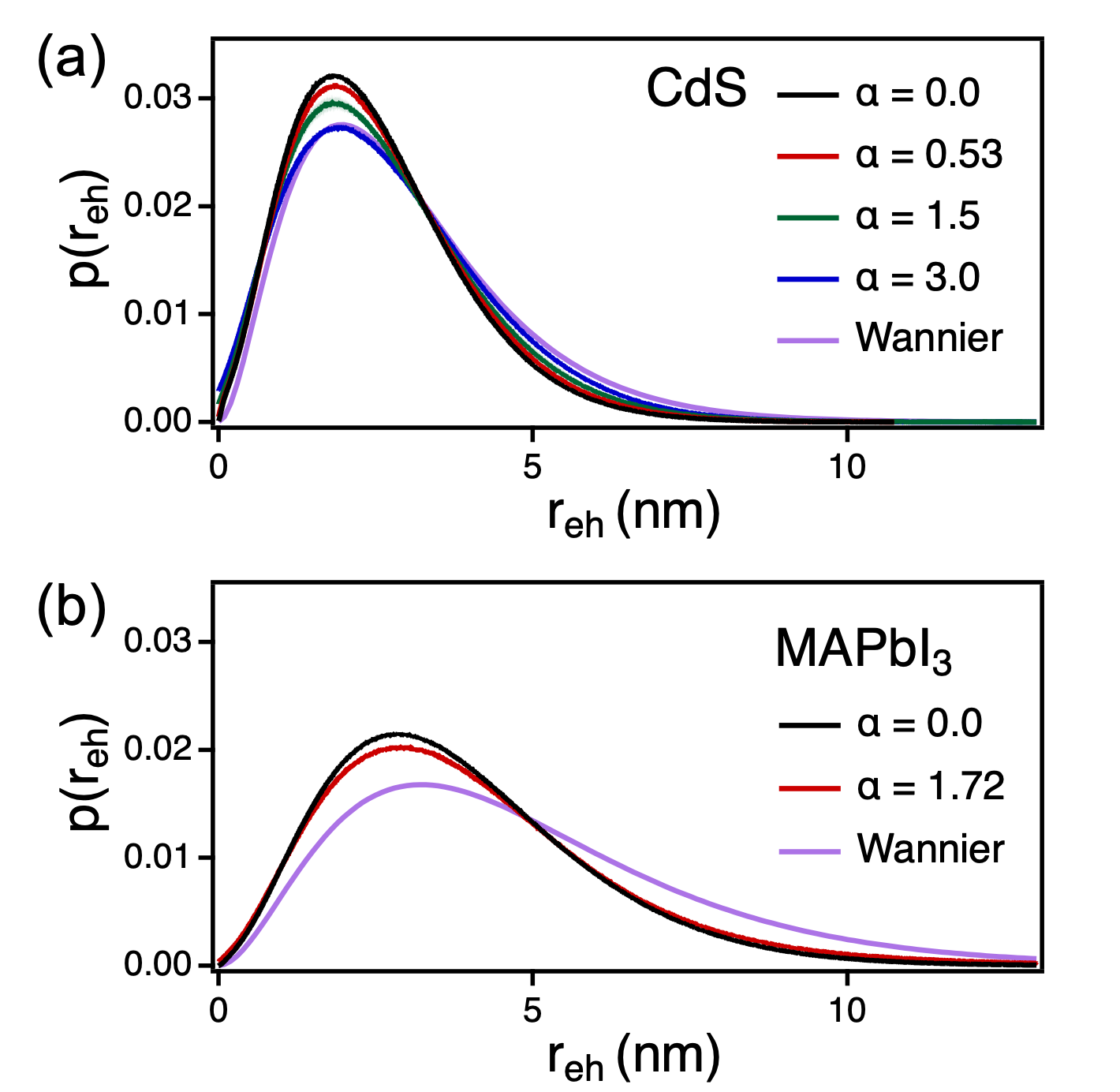}
\caption{ Electron-hole charge density distributions. Charge density distributions for (a) CdS and (b) MAPbI$_3$ for a variety of electron-phonon coupling strength $\alpha$ evaluated at $T=$50 K, compared to the Wannier model. Error bars are shown as shaded regions.}
\label{monopr}
\end{figure}

In addition to the decreased likelihood of finding electron and hole pairs together, with increasing $\alpha$ the charge density distribution is broadened. 
Shown in Fig.~\ref{monopr} are the probability distributions of the distance between electron and hole beads, $r_{eh}=|\mathbf{x}_e-\mathbf{x}_h|$, at each strength of interaction with phonons. In both materials, the stronger the phonon interaction is, the larger the average bead-bead distance becomes, implying that phonons make an effective electron-hole interaction weaker through screening. The comparison between results from path integral simulation with the probability distribution predicted from Wannier-Mott exciton using hydrogen model shown in purple lines in Fig.~\ref{monopr} implies the importance of capturing the fluctuations of quasiparticles at room temperature.  

The ratio of path partition functions $\mathcal{Z}_c/\mathcal{Z}_0$ for the bimolecular rate constant can be computed and extrapolated to the large $n$ limit in order to extract a converged overlap element, as shown in App.~\ref{App3}. The calculated bimolecular recombination rates for both materials with typical charge density $\rho_e = 10^{17} \mathrm{cm}^{-3}$ are summarized in Table \ref{ratesummary} as a function of $\alpha$. The rates are found for both materials to decrease significantly over the range of electron-phonon coupling strength considered. MAPbI$_3$ is found to have a longer charge carrier lifetime than the CdS, which is due to enhanced screening of the former. 

%%%%%%%%%%%%%%%%%%%%%%%%%%%%%%%%%%%%%%%%%%%%%%%%%%%%%%%%%

\subsection{Total recombination rate}

Combining trap-assisted and bimolecular rates, the total recombination rates defined in Eq.~\ref{ratelaw} are summarized in Table \ref{ratesummary} with a typical charge carrier density $\rho_e = 10^{17} \mathrm{cm}^{-3}$ and the defect density $\rho_{\mathrm{d}} = 10^{18} \mathrm{cm}^{-3}$ for both materials. 
We find that the radiative recombination is predominantly determined by bimolecular process. In both materials, electron-phonon coupling generally decreases recombination rate, resulting in the increase in the lifetime of charge carriers. The value obtained for CdS is in very good agreement with that observed from photoluminscence lifetime measurements on large spherical nanocrystals, 13 ns, but underestimates the lifetime reported for MAPI$_3$ by a factor of 2, 70-100 ns.\cite{herz2016} 
\textcolor{black}{The latter disagreement can be attributed to the neglect of anharmonic effects accounted by the $\mathbf{k}$-dependence of correlation function in the optical mode which can been considered previously\cite{park2022} and results in an effective electron-hole repulsion that has not been considered here.}

\section{Conclusions}

In summary, we have shown how a path integral approach can be used to study excitonic properties in the presence of dynamical phonons. We have presented ways to compute the renormalization of the binding energy and recombination rate, and validated these results in limiting regimes. While we have considered a simple model for an exciton in a polar lattice as being coupled through a Fr\"ohlich interaction with a single optical mode, the influence functional approach employed is general and can easily be extended to many modes, arbitrary linear coupling forms, and in principle parameterized through \emph{ab initio} methods. Additionally, while we have considered phonon effects on a single exciton in this work, this can be applied to multiple excitons and further combined with confining potentials.

%%%%%%%%%%%%%%%%%%%%%%%%%%%%%%%%%%%%%%%%%%%%%%%%%%%%%%%%%
%%%%%%%%%%%%%%%%%%%%%%%%%%%%%%%%%%%%%%%%%%%%%%%%%%%%%%%%%
%%%%%%%%%%%%%%%%%%%%%%%%%%%%%%%%%%%%%%%%%%%%%%%%%%%%%%%%%
\vspace{1mm}
\begin{acknowledgments}
This work was supported by the U.S. Department of Energy, Office of Science, Office of Basic Energy Sciences, Materials Sciences and Engineering Division under Contract No. DEAC02-05-CH11231 within the Physical Chemistry of Inorganic Nanostructures Program (No. KC3103). This research used resources of the National Energy Research Scientific Computing Center (NERSC), a U.S. Department of Energy Office of Science User Facility. Y.P. acknowledges the Kwanjeong Educational Foundation. D.T.L. acknowledges the Alfred P. Sloan Foundation.
\end{acknowledgments}

\appendix

\section{Generalized influence functional}
\label{App1}
In the main text we have focused on a simplified model whereby the electron and hole are coupled to a single optical phonon through a linear Fr\"ohlich coupling. For more complex lattice models where multiple dispersive modes are relevant, the influence functional formalism employed can be generalized. For many modes, the classical Hamiltonian entering into the path action can be written as
\begin{equation}
\mathcal{H}_\mathrm{ph} = \frac{1}{2 \beta \hbar} \int_{\tau=0}^{\beta \hbar}  \sum_{\mathbf{k}} (\dot{\mathbf{q}}_{\mathbf{k},\tau}^2 + \omega_{\mathbf{k}}^2 {\mathbf{q}}_{\mathbf{k},\tau}^2)
\end{equation}
where $\mathbf{q}_{\mathbf{k},\tau}$ is the classical displacement of the $\mathbf{k}$ mode at imaginary time $\tau$, and $\omega_{\mathbf{k}}$ is its corresponding frequency. For a generalized linear coupling between the charge density and the lattice of the form
\begin{equation}
\mathcal{H}_\mathrm{int} 
=\frac{1}{\beta \hbar} \int_{\tau=0}^{\beta \hbar} \sum_{\mathbf{k}} \ {\mathbf{q}}_{\mathbf{k},\tau} \ \frac{C_{e,\mathbf{k}} e^{i {\mathbf{k}} \cdot {\mathbf{x}}_{e,\tau}} - C_{h,\mathbf{k}} e^{i {\mathbf{k}} \cdot {\mathbf{x}}_{h,\tau}}}{\mathbf{k}}
\end{equation}
where $C_{e/h,\mathbf{k}}$ are generalized coupling coefficients, the phonons can still be integrated out. This yields an effective potential between the electron and hole of the form
\begin{eqnarray}
\mathcal{H}_{\mathrm{eff,\tau}}^{ij} &&=\\
&& - \frac{\sigma_{ij}}{2 \beta \hbar} \sum_{\mathbf{k}} \int_{\tau' = 0}^{\beta \hbar } C_{i,\mathbf{k}} C^*_{j,\mathbf{k}} \frac{e^{- i \mathbf{k} \cdot |\mathbf{x}_{i,\tau} - \mathbf{x}_{j,\tau'}|}}{k^2} \chi_\mathbf{k}(\tau-\tau') \nonumber
\end{eqnarray}
where $\chi_\mathbf{k}(\tau-\tau')=\langle \mathbf{q}_{\mathbf{k}}(0) \mathbf{q}_{\mathbf{k}}(\tau-\tau') \rangle$ is the imaginary time correlation function of mode $\mathbf{q}_{\mathbf{k}}$. In Fourier space, the correlation function is given by
\begin{equation}
\chi_\mathbf{k}(\omega)= \frac{1}{\omega^2+\omega_{\mathbf{k}}^2}
\end{equation}
a sum of poles. In the classical limit, $\beta \hbar \omega_\mathbf{k} \rightarrow 0$, this influence functional returns a Coulomb potential screened by a wavevector dependent dielectric susceptibility.\cite{park2022} 

\begin {figure}[t]
\centering\includegraphics [width=8.3cm] {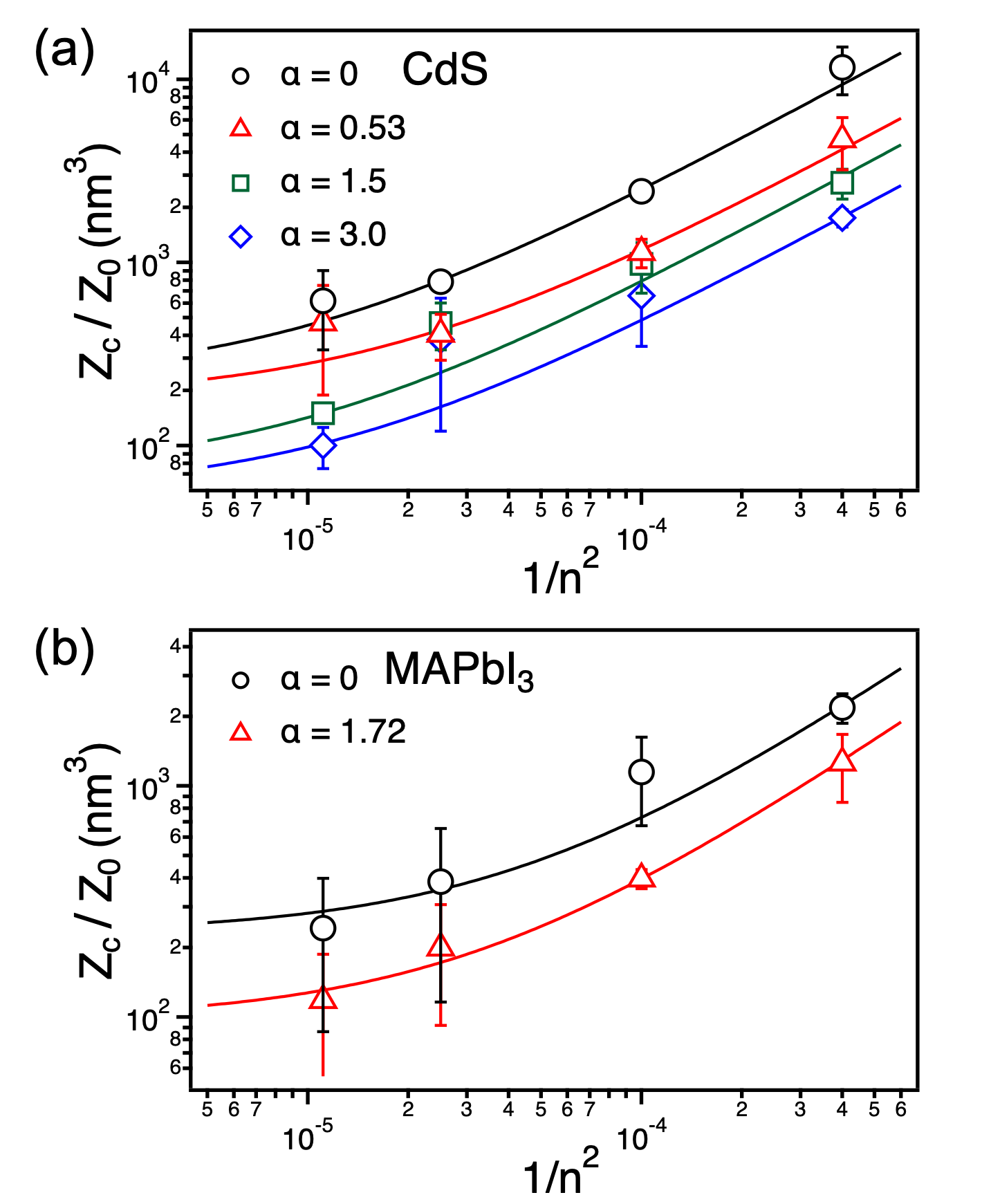}
\caption{Relative path partition functions for the radiative rate calculation. Convergence of the overlap for (a) CdS and (b) MAPbI$_3$. The values of $\alpha$ considered are $\{0,0.53,1.5,3\}$ for CdS and $\{0,1.72\}$ for MAPbI$_3$. Solid lines have the form $a+b/n^2$ where $a$ and $b$ are positive constants.}
\label{rate}
\end{figure}

\section{Pollmann-Buttner theory} 
\label{App2}

In order to compare our exciton binding energies to Pollman-Buttner theory,\cite{pbtheory} we parameterize an effective potential of the form
\begin{eqnarray}
V_\mathrm{eff}(r) &=& -\frac{e^2}{4\pi \varepsilon_s r}  \\
 &&-\frac{e^2}{4\pi  \varepsilon^* \Delta m r} \left (m_h e^{-r/R_h} - m_e e^{r/R_e} \right ) \nonumber
\end{eqnarray}
\textcolor{black}{where $\Delta m = m_h-m_e$, $1/\varepsilon^* = 1/\varepsilon_r - 1/\varepsilon_s$, $R_{e/h} = \sqrt{\hbar /2 m_{e/h} \omega}$ with a static dielectric constant $\varepsilon_s$, which is taken to be 8.9 and 24.1 for CdS and MAPbI$_3$.} We solve the time independent Schrodinger equation using this potential in the relative coordinate system for the electron and hole, $\mathbf{r}_{eh}$. This takes the form
\begin{equation}
-\frac{\hbar^2}{2 \mu }\nabla^2 \phi_n(\mathbf{r}_{eh}) + V_\mathrm{eff}(r_{eh}) \phi_n(\mathbf{r}_{eh}) = E_n \phi_n(\mathbf{r}_{eh})
\end{equation}
where $E_n$ and  $\phi_n(\mathbf{r}_{eh})$ are the associated eigenvalues and eigenvectors.  Solving this equation for the ground state with zero angular momentum simplifies this to putting the equation on a \textcolor{black}{real space grid on $r_{eh}$}, which yields the exciton binding energies reported in the main text.

\section{Overlap density convergence}
\label{App3}
The bimolecular rate constant $k_{\mathrm{b}}$ is determined by the ratio of partition functions $\mathcal{Z}_c/\mathcal{Z}_0$.
The error from the path integral scales by the number of discretization factor as $n^{-2}$,  so the convergence of the radiative rate can be extrapolated by fitting the ratios of partition functions at finite $n$  as a function of $n^{-2}$. Extrapolations for CdS and MAPbI$_3$ are plotted in Fig.~\ref{rate} (a) and (b) under different coupling strengths. Since the path integral formalism becomes exact in the limit of large number beads, the rate constant is defined as the value extrapolated to $1/n^2 \rightarrow 0$ limit. 

\section*{References}
%%%%%%%%%%%%%%%%%%%%%%%%%%%%%%%%%%%%%%%%%%%%%%%%%%%%%%%%%
%%%%%%%%%%%%%%%%%%%%%%%%%%%%%%%%%%%%%%%%%%%%%%%%%%%%%%%%%
\bibliography{main}
%%%%%%%%%%%%%%%%%%%%%%%%%%%%%%%%%%%%%%%%%%%%%%%%%%%%%%%%%
\end{document}